\documentclass[12pt,a4paper]{article}
\usepackage{amsmath}	
\usepackage{amssymb}
\usepackage{color}
\usepackage{caption}
\usepackage{graphicx}
\usepackage{subcaption}
\usepackage{hyperref}

\def\DIANA{\texttt{DIANA}}
\def\MATAD{\texttt{MATAD} }
\def\BAMBA{\texttt{BAMBA} }
\def\COLOR{\texttt{COLOR} }

\def\lam{\ensuremath{\hat\lambda}}
\newcommand{\ilam}{\ensuremath {\frac{1}{\lam}}}

\newcommand{\ala}{\ensuremath {a_1}}
\newcommand{\alb}{\ensuremath {a_2}}
\newcommand{\alc}{\ensuremath {a_s}}

\newcommand{\MS}{{\ensuremath{\overline{\mathrm{MS}}}}}
\def\z#1{{\zeta_{#1}}}
\def\LYukawa{\ensuremath{\mathcal{L}_{\mathrm{Yukawa}}}}
\def\LH{\ensuremath{\mathcal{L}_{\mathrm{H}}}}
\DeclareMathOperator{\tr}{tr}

\newcommand{\YtD}{\ensuremath {\mathcal{Y}_{d}}}
\newcommand{\YtU}{\ensuremath {\mathcal{Y}_{u}}}
\newcommand{\YtL}{\ensuremath {\mathcal{Y}_{l}}}

\newcommand{\YtDD}{\ensuremath {\mathcal{Y}_{dd}}}
\newcommand{\YtUU}{\ensuremath {\mathcal{Y}_{uu}}}
\newcommand{\YtLL}{\ensuremath {\mathcal{Y}_{ll}}}
\newcommand{\YtUD}{\ensuremath {\mathcal{Y}_{ud}}}

\newcommand{\YtDDD}{\ensuremath {\mathcal{Y}_{ddd}}}
\newcommand{\YtUUD}{\ensuremath {\mathcal{Y}_{uud}}}
\newcommand{\YtUDD}{\ensuremath {\mathcal{Y}_{udd}}}
\newcommand{\YtUUU}{\ensuremath {\mathcal{Y}_{uuu}}}
\newcommand{\YtLLL}{\ensuremath {\mathcal{Y}_{lll}}}

\newcommand{\YtUUUU}{\ensuremath {\mathcal{Y}_{uuuu}}}
\newcommand{\YtUUUD}{\ensuremath {\mathcal{Y}_{uuud}}}
\newcommand{\YtUUDD}{\ensuremath {\mathcal{Y}_{uudd}}}
\newcommand{\YtUDUD}{\ensuremath {\mathcal{Y}_{udud}}}
\newcommand{\YtUDDD}{\ensuremath {\mathcal{Y}_{uddd}}}
\newcommand{\YtDDDD}{\ensuremath {\mathcal{Y}_{dddd}}}
\newcommand{\YtLLLL}{\ensuremath {\mathcal{Y}_{llll}}}

\newcommand{\NGen}{\ensuremath {n_G}}

\def \eps {\epsilon}

\begin{document}

\begingroup\raggedleft\footnotesize\ttfamily
HU-Mathematik-2013-20\\
HU-EP-13/53\\

\vspace{15mm}
\endgroup

\begin{center}
{\Large{\bf
Three-loop Higgs self-coupling beta-function \\ in the Standard Model \\[3mm] with complex Yukawa matrices
}}
\vspace{15mm}

{\sc
A.~V.~Bednyakov${}^1$,
A.~F.~Pikelner${}^1$
and V.~N.~Velizhanin${}^{2,3}$}\\[5mm]

${}^1${\it
Joint Institute for Nuclear Research,\\
 141980 Dubna, Russia}

\vspace{5mm}

${}^2${\it
\mbox{Institut f{\"u}r  Mathematik und Institut f{\"u}r Physik, Humboldt-Universit{\"a}t zu Berlin,}\\
IRIS Adlershof, Zum Gro\ss en Windkanal 6, 12489 Berlin, Germany
}

\vspace{3mm}

${}^3${\it
Theoretical Physics Division, Petersburg Nuclear Physics Institute,\\
  Orlova Roscha, Gatchina, 188300 St.~Petersburg, Russia}

\vspace{3mm}

\vspace{15mm}

\textbf{Abstract}\\[2mm]
\end{center}

\noindent{Three-loop renormalization group equations for the Higgs self-coupling and Higgs mass parameter are recalculated in the case of complex Yukawa matrices
which encompass the general flavour structure of the Standard Model. 
In addition, the anomalous dimensions for both the quantum Higgs field and 
its vacuum expectation value are presented in the \MS-scheme. 
A numerical study of the latter quantities is carried out for a certain set 
of initial parameters.
}
\newpage

\setcounter{page}{1}

The discovery of the Higgs boson \cite{Aad:2012tfa,Chatrchyan:2012ufa} confirms the fact that the Standard Model turns out to be a perfect model describing physics at the electroweak scale. 
In spite of all attempts to find something beyond the SM, no stringent evidences of new particles were found.

Recent analyses~\cite{Bezrukov:2012sa,Alekhin:2012py,Degrassi:2012ry,Buttazzo:2013uya} based on three-loop renormalization group equations \cite{Mihaila:2012fm,Bednyakov:2012en,Chetyrkin:2013wya} 
demonstrated that the SM can be extrapolated up to very high scales without the necessity to introduce additional degrees of freedom.

Unfortunately, current experimental uncertainty in the strong coupling constant and the top quark mass do not allow us 
to make an accurate prediction whether the SM vacuum is stable only up to $\mathcal{O}(10^{10})$~GeV or up to the Plank scale. 
It is not surprising that in the above-mentioned studies focused on vacuum stability the flavour structure of the SM was neglected. 

In this work, we extend our recent results on Higgs potential parameters to the case of general Yukawa matrices.
This kind of result can be important not only in precise studies of vacuum stability, but also in an analysis of different flavour patterns 
(see, e.g., a review \cite{Fritzsch:1999ee})), which can again originate from some New Physics. 

The corresponding two-loop expressions \cite{Luo:2002ey} can be deduced from the general results of Refs.~\cite{Machacek:1983tz,Machacek:1983fi,Machacek:1984zw,Luo:2002ti}. 
The three-loop gauge-coupling beta-functions with the full flavour structure were calculated for the first time in Ref.~\cite{Mihaila:2012pz} and confirmed later
by our group \cite{Bednyakov:2012rb}. 
It is worth mentioning that contrary to these results, which were found from the corrections in the SM with diagonal Yukawa couplings by certain substitutions, 
	the expressions presented in this paper are obtained 
	by direct calculation of Feynman diagrams with explicit flavour indices.

	For this kind of calculation the Feynman rules for \DIANA~\cite{Tentyukov:1999is}, which were used in our previous studies, were appropriately rewritten and a simple
	routine dealing with explicit flavour indices was developed.
In order to validate our codes, we also recalculated the results for the gauge coupling beta-functions, thus confirming the expressions given in Refs~\cite{Mihaila:2012pz,Bednyakov:2012rb}. 

The calculation is carried out in an almost automatic way with the help of the infra-red rearrangement (IRR)~\cite{Vladimirov:1979zm}
procedure implemented in our codes. 
We start with the Lagrangian of the unbroken SM with the full flavour structure given in our previous paper~\cite{Bednyakov:2012rb}.
For the reader's convenience we present here the terms describing the fermion-Higgs interactions and the Higgs field self-interaction
	\begin{eqnarray}
	  \LYukawa &=& - \bigg(
	Y^{ij}_u (Q^L_i \Phi^c) u^R_j
	+ Y^{ij}_d (Q^L_i \Phi) d^R_j  + Y^{ij}_l (L^L_i \Phi) l^R_j
	+ \mathrm{h.c.} \bigg),
	  \label{eq:yukawa_lag}\\[3mm]
          \LH & = & \left( D_\mu \Phi \right)^\dagger  \left( D_\mu \Phi \right) - V_H(\Phi)\,,
	  \label{eq:higgs_lag} \\[1mm]
	  V_H(\Phi) & = &   m^2 \Phi^\dagger \Phi + \lambda \left( \Phi^\dagger \Phi \right)^2,\qquad \Phi^\dagger \Phi =  \frac{ h^2 + \chi^2}{2} +  \phi^+ \phi^- 
	  \label{eq:higgs_pot}
        \end{eqnarray}
	Here $\lambda$ and $Y_{u,d,l}$ denote the Higgs quartic
	and  Yukawa matrices, respectively. The left-handed quark and lepton SU(2) doublets, $Q^L_i$, and $L^L_i$, carry flavour indices $i=1,2,3$.  
	The same is true for the SU(2) singlets corresponding to the right-handed SM fermions  
	$u_i^R$, $d_i^R$, and $L^R_i$.
The Higgs doublet $\Phi$ with hypercharge $Y_W = 1$ is decomposed in terms of the component fields:
\begin{equation}
	\Phi =
	\left(
	\begin{array}{c}
		\phi^+(x) \\ \frac{1}{\sqrt 2} \left( h + i \chi \right)
		\end{array}
	\right),
	\qquad
	\Phi^c = i\sigma^2 \Phi^\dagger =
	\left(
	\begin{array}{c}
		\frac{1}{\sqrt 2} \left( h - i \chi \right) \\
		-\phi^-
		\end{array}
	\right).
	\label{eq:Phi_def}
\end{equation}
	The charge-conjugated Higgs doublet $\Phi^c$ has $Y_W=-1$ and enters into the Yukawa interactions of the right-handed up-type quarks.
We neglect the Higgs mass parameter in the Lagrangian since the corresponding anomalous dimension can be found from the $\MS$-renormalization 
constant of the $|\Phi|^2$ operator (see, e.g., \cite{Chetyrkin:2012rz,Bednyakov:2013eba}).
	

The utilized IRR prescription consists of the introduction of an auxiliary mass parameter $M$ in every propagator
and the subsequent expansion in external momenta. The details on this technique can be found in Refs.~\cite{Misiak:1994zw,Chetyrkin:1997fm}. 
The resulting fully massive vacuum integrals can be easily evaluated by means of the \MATAD~package~\cite{Steinhauser:2000ry} or \BAMBA~code developed by V.N.~Velizhanin.
For the color algebra the FORM package \COLOR~\cite{vanRitbergen:1998pn} is utilized. 
It is worth mentioning that we recalculated all needed two-loop counter-terms, for both the SM parameters and the auxiliary boson masses. 

In order to find the renormalization constants for $\lambda$ we consider symmetric four-point Green functions with external Higgs particles $h$. 
A special script which takes into account the permutation symmetry of external lines, allows us to substantially reduce the number of calculated three-loop diagrams 
(from about 8 million to about 600 thousand). It is worth mentioning that the number of diagrams, which 
has to be evaluated, can be further reduced with the help of the \texttt{graph\_state} library \cite{graphstate}
(by about 200 thousand in the considered case). 
The latter allows one to find isomorphic Feynman diagrams by using the generalization of graph 
labelling and ordering algorithm\footnote{The generalization  also takes into account fields on internal lines.}
proposed in \cite{Nickel} (Nickel index).

As in our previous paper \cite{Bednyakov:2013eba} the anomalous dimension of the Higgs mass parameter $m^2$ is inferred from 
a certain set of Feynman diagrams contributing to the scalar four-point Green function with two neutral and two charged external Higgs bosons. In all diagrams from this set both lines associated with external charged particles are connected to a single quartic vertex that mimics the insertion of the $|\Phi|^2$ operator. 

From the corresponding renormalization constants $Z_{hhhh}$ and $Z_{hh[\phi^+\phi^-]}$ we 
obtain ($\lam\equiv\lambda/(16\pi^2)$)
\begin{equation}
	Z_{\lam} = \frac{Z_{hhhh}}{Z_h^2} 
	\qquad
	Z_{m^2} = \frac{Z_{hh[\phi^+\phi^-]}}{Z_{h}}
	\label{eq:lambda_mass_RC}
\end{equation}
where $Z_h$ is nothing else but the renormalization constant for the Higgs propagator\footnote{Due to unbroken SU(2) invariance all the fields from the Higgs doublet
	have the same renormalization constant $Z^{1/2}_h$.}, and
$Z_{\lam}$, $Z_{m^2}$ enter into the relations between the bare parameters $\lam_{\mathrm{Bare}}$, $m^2_{\mathrm{Bare}}$ and the corresponding renormalized ones
\begin{eqnarray}
	\lam_{\mathrm{Bare}}\mu^{-2\epsilon} & = &  Z_{\lam} \lam = 
	\lam 
	+ \sum_{l=1}^\infty\sum_{n=1}^l \frac{c_{\lam}^{(l,n)}}{\epsilon^n} 
	 	\label{eq:Z_lambda} \\
	m^2_{\mathrm{Bare}} & = & Z_{m^2} m^2 
	= m^2 \left( 1  
	+ \sum_{l=1}^\infty\sum_{n=1}^l \frac{c_{m^2}^{(l,n)}}{\epsilon^n}\right). 
	\label{eq:Z_ms}
\end{eqnarray}
	Here $\mu$ is the \MS~renormalization scale, $\epsilon = (4-D)/2$ is the parameter of dimensional regularization, and $c_{\lambda, m^2}^{(l,n)}$ denotes the $l$-loop contribution to the coefficient of $1/\eps^n$
	in the considered renormalization constants.

The required renormalization group coefficients are extracted from the single pole in $\eps$ with the help of the following formulae: 
\begin{equation}
	\beta_{\lam} = \frac{d \lam(\mu,\epsilon)}{d \ln \mu^2}\bigg|_{\epsilon=0} = 
	\sum_{l=1}^{\infty} l \cdot c_{\lam}^{(l,1)},
	\quad
	\gamma_{m^2} = \frac{d \ln m^2(\mu,\epsilon)}{d \ln \mu^2}\bigg|_{\epsilon=0} = 
	\sum_{l=1}^{\infty} l \cdot c_{m^2}^{(l,1)},
\end{equation}

The explicit expressions for $\beta_{\lam}$ and $\gamma_{m^2}$ can be found in ancillary files of the arXiv version of the paper. 
The results depend on traces of different combinations of the Yukawa matrices which we list here 
($f = u,d,l$)	for convenience
\begin{eqnarray}
	\mathcal{Y}_{f} \equiv  \frac{\tr Y_f Y_f^\dagger}{16\pi^2},
	& \qquad &	
	\mathcal{Y}_{ff} \equiv  \frac{\tr Y_f Y_f^\dagger Y_f Y_f^\dagger}{(16\pi^2)^2}, \nonumber\\
	\mathcal{Y}_{fff} \equiv  \frac{\tr Y_f Y_f^\dagger Y_f Y_f^\dagger Y_f Y_f^\dagger}{(16\pi^2)^3},
	& \qquad &	
	\mathcal{Y}_{ffff} \equiv  \frac{\tr Y_f Y_f^\dagger Y_f Y_f^\dagger Y_f Y_f^\dagger Y_f Y_f^\dagger}{(16\pi^2)^4} \nonumber
\end{eqnarray}
\begin{eqnarray}
	  \YtUD =  \frac{\tr Y_u Y_u^\dagger Y_d Y_d^\dagger}{(16\pi^2)^2},
	& \qquad &
	  \YtUDD =  \frac{\tr Y_u Y_u^\dagger Y_d Y_d^\dagger Y_d Y_d^\dagger}{(16\pi^2)^3}, \nonumber\\
	  \YtUUD =  \frac{\tr Y_u Y_u^\dagger Y_u Y_u^\dagger Y_d Y_d^\dagger}{(16\pi^2)^3},
	& \qquad &
	  \YtUUUD =  \frac{\tr Y_u Y_u^\dagger Y_u Y_u^\dagger Y_u Y_u^\dagger Y_d Y_d^\dagger}{(16\pi^2)^4}, \nonumber\\
	  \YtUUDD =  \frac{\tr Y_u Y_u^\dagger Y_u Y_u^\dagger Y_d Y_d^\dagger Y_d Y_d^\dagger}{(16\pi^2)^4},
	& \qquad &
	  \YtUDUD =  \frac{\tr Y_u Y_u^\dagger Y_d Y_d^\dagger Y_u Y_u^\dagger Y_d Y_d^\dagger }{(16\pi^2)^4}, 
	  \nonumber
\end{eqnarray}
\begin{eqnarray}
	  \YtUDDD =  \frac{\tr Y_u Y_u^\dagger Y_d Y_d^\dagger Y_d Y_d^\dagger Y_d Y_d^\dagger}{(16\pi^2)^4}.
\end{eqnarray}
In these expressions the product $Y_f Y_f^\dagger$ corresponds to the propagation of the right-handed fermion $f$. Since there is no
right-handed flavour-changing current coupled to a SM gauge field, the expressions of the form $Y_{f'} Y_f^\dagger$ with $f\neq f'$ do not appear in the
results.

If one neglects the mixing between generations together with the Yukawa couplings of the first two
fermion families, one obtains the known expressions \cite{Bednyakov:2012en}.
To save space, we do not show the results for $\lam$ and $m^2$ themselves but present here an interesting combination of 
these quantities which can be associated with the three-loop anomalous dimension 
$\gamma_v$ of the ``tree-level'' vacuum expectation value defined by the expression
$v(\mu) = \sqrt{-\frac{m^2(\mu)}{\lambda(\mu)}}$ (see, e.g., Ref.~\cite{Jegerlehner:2012kn}).
From this definition one can deduce that
\begin{equation}
	\gamma_v = \frac{1}{2} \left( \gamma_{m^2} - \frac{\beta_{\lam}}{\lam} \right) = \gamma_v^{(1)} + \gamma_v^{(2)} + \gamma_v^{(3)} + \dots,
\end{equation}
	with $\gamma_v^{(l)}$ being the $l$-loop contribution given by the following expressions 

{\allowdisplaybreaks
\begin{eqnarray}
	\gamma_v^{(1)} && = 
\ilam \left(\frac{3 (\YtDD+\YtUU)}{2}+\frac{\YtLL}{2}\right)
-\frac{9}{80} \frac{\ala \alb}{\lam}
-\frac{27}{800} \frac{\ala^2}{\lam}
-\frac{9}{32}\frac{\alb^2 }{\lam}
\nonumber\\
&&
-3 \left(\lam+\frac{\YtD+\YtU}{2}\right)
-\frac{\YtL}{2}
+\frac{9 \ala}{40}
+\frac{9 \alb}{8},
\label{eq:anom_v_1}
\\
	\gamma_v^{(2)} && = 
\ilam \left(\frac{3}{2} (\YtUDD+\YtUUD)-\frac{15 (\YtDDD+\YtUUU)}{2}-\frac{5 \YtLLL}{2}\right)
-10 \alc (\YtD+\YtU)
\nonumber\\
&&
-\frac{\ala \alb }{\lam} \left(\frac{27 \YtD}{40}+\frac{33 \YtL}{40}+\frac{63 \YtU}{40}\right)
-\ala \left(\frac{9 \lam}{5}+\frac{5 \YtD}{16}+\frac{15 \YtL}{16}+\frac{17 \YtU}{16}\right)
\nonumber\\
&&
+\frac{\ala^2 }{\lam} \left(-\frac{9 \YtD}{80}+\frac{9 \YtL}{16}+\frac{171 \YtU}{400}\right)
+\frac{\ala }{\lam} \left(-\frac{\YtDD}{5}+\frac{3 \YtLL}{5}+\frac{2 \YtUU}{5}\right)
+\frac{21 \YtUD}{4}
\nonumber\\
&&
-\alb \left(9 \lam+\frac{45 (\YtD+\YtU)}{16}+\frac{15 \YtL}{16}\right)
+\frac{\alb^2 }{\lam} \left(\frac{9 (\YtD+\YtU)}{16}+\frac{3 \YtL}{16}\right)
-\frac{7 \YtLL}{8}
\nonumber\\
&&
+\frac{\ala^2 \alb }{\lam} \left(\frac{\NGen}{5}+\frac{717}{1600}\right)
+\frac{\ala \alb^2 }{\lam} \left(\frac{\NGen}{5}+\frac{97}{320}\right)
+\frac{\ala^3 }{\lam} \left(\frac{3 \NGen}{25}+\frac{531}{8000}\right)
\nonumber\\
&&
+\ala^2 \left(-\frac{\NGen}{4}-\frac{903}{1600}\right)
+\frac{\alb^3 }{\lam} \left(\NGen-\frac{497}{64}\right)
+\alb^2 \left(\frac{241}{64}-\frac{5 \NGen}{4}\right)
+63 \lam^2
\nonumber\\
&&
+\left(\frac{8 \alc }{\lam} -\frac{21}{8}\right) (\YtDD+\YtUU)
-\frac{189 \ala \alb}{160}
+18 \lam (\YtD+\YtU)
+6 \lam \YtL,
\label{eq:anom_v_2}
\\
	\gamma_v^{(3)} && = 
\frac{\ala^3 \alb }{\lam} \left(\frac{\NGen^2}{9}+\NGen \left(\frac{18001}{24000}-\frac{183 \z3}{250}\right)-\frac{81 \z3}{320}+\frac{29779}{64000}\right)
\nonumber\\
&&
+\frac{\ala^2 \alb^2 }{\lam} \left(\frac{\NGen^2}{9}-\NGen \left(\frac{63 \z3}{50}+\frac{149}{3600}\right)-\frac{7857 \z3}{3200}+\frac{64693}{19200}\right)
\nonumber\\
&&
+\frac{\ala^4 }{\lam} \left(\frac{\NGen^2}{10}+\NGen \left(\frac{12441}{16000}-\frac{171 \z3}{250}\right)-\frac{8019 \z3}{160000}+\frac{12321}{256000}\right)
\nonumber\\
&&
+\frac{\alb^4 }{\lam} \left(\frac{5 \NGen^2}{6}+\NGen \left(\frac{45 \z3}{2}+\frac{14749}{384}\right)+\frac{2781 \z3}{256}-\frac{982291}{6144}\right)
\nonumber\\
&&
+\frac{\ala \alc }{\lam} \left(-\frac{68 \YtDD \z3}{5}+\frac{641 \YtDD}{60}+\frac{28 \YtUU \z3}{5}-\frac{931 \YtUU}{60}\right)
-\frac{6 }{\lam} \YtLL \YtUD
\nonumber\\
&&
+\ala^3 \left(-\frac{7 \NGen^2}{18}+\NGen \left(\frac{57 \z3}{50}-\frac{1523}{600}\right)-\frac{1863 \z3}{4000}-\frac{9323}{4000}\right)
+\frac{9 \ala \YtL^2}{40}
\nonumber\\
&&
+\frac{\alb \alc }{\lam} \left(-12 \z3 (\YtDD+\YtUU)+\frac{31 (\YtDD+\YtUU)}{4}-48 \YtUD \z3+4 \YtUD\right)
\nonumber\\
&&
+\alb^3 \left(-\frac{35 \NGen^2}{18}-\NGen \left(\frac{45 \z3}{2}+\frac{4163}{144}\right)-\frac{3807 \z3}{32}+\frac{53563}{1152}\right)
\nonumber\\
&&
+\frac{\ala \alb \alc }{\lam} \left(\frac{54}{5} \z3 (\YtD+\YtU)-\frac{699 \YtD}{40}-\frac{747 \YtU}{40}\right)
-90 \alb \lam^2 \z3
\nonumber\\
&&
+\frac{\ala^2 \alc }{\lam} \left(\frac{81}{25} \z3 (\YtD+\YtU)-\frac{2049 \YtD}{400}-\frac{1761 \YtU}{400}\right)
+324 \lam \YtUD \z3
\nonumber\\
&&
+\ala \alc \left(-6 \YtD \z3+\frac{991 \YtD}{120}-\frac{102 \YtU \z3}{5}+\frac{2419 \YtU}{120}\right)
+162 \lam \YtD \YtU
\nonumber\\
&&
+\frac{\ala \alb^3 }{\lam} \left(\frac{\NGen^2}{9}+\frac{8341 \NGen}{2880}+\frac{243 \z3}{64}+\frac{54053}{11520}\right)
+\frac{\alc }{\lam} \YtUUD
+\frac{\alc }{\lam} \YtUDD
\nonumber\\
&&
+\ala^2 \alb \left(\NGen \left(\frac{27 \z3}{50}-\frac{243}{80}\right)-\frac{1809 \z3}{800}-\frac{25767}{1600}\right)
+\frac{9 \alb \YtL^2}{8}
-\frac{41 \alc \YtUD}{2}
\nonumber\\
&&
+\frac{\ala \alb^2 }{\lam} \NGen \left(\frac{3 \YtL}{20}-3 \left(\frac{\YtD}{4}+\frac{\YtU}{20}\right)\right)
+\frac{\ala^2 \alb }{\lam} \NGen \left(\frac{3 \YtL}{20}-3 \left(\frac{\YtD}{4}+\frac{\YtU}{20}\right)\right)
\nonumber\\
&&
+\ala \alb^2 \left(\NGen \left(\frac{9 \z3}{10}-\frac{27}{8}\right)+\frac{279 \z3}{160}-\frac{3849}{128}\right)
-\frac{21 \YtL \YtUD}{4}
-90 \lam \YtLL \z3
\nonumber\\
&&
+\ala^2 \NGen \left(\frac{129 \lam}{20}+\frac{31 \YtD}{80}+\frac{117 \YtL}{80}+\frac{127 \YtU}{80}\right)
+24 \alc \YtUD \z3
+\frac{3145 \YtLLL}{32}
\nonumber\\
&&
+\alc^2 \left(16 \NGen (\YtD+\YtU)+12 \z3 (\YtD+\YtU)-\frac{455 (\YtD+\YtU)}{3}\right)
\nonumber\\
&&
-\frac{\ala^3 }{\lam} \NGen \left(\frac{57 \YtD}{100}+\frac{99 \YtL}{100}+\frac{129 \YtU}{100}\right)
-1008 \lam^3 \z3
+9 \lam \YtL^2
-28 \YtL \YtLL
\nonumber\\
&&
+\frac{\alb^2 }{\lam} \NGen \left(\frac{39 (\YtDD+\YtUU)}{8}+\frac{13 \YtLL}{8}+6 \YtUD\right)
+\ala^2 \NGen
+\frac{\ala^2 }{\lam}
+\ala \alb^2
\nonumber\\
&&
+\frac{\ala^2 }{\lam} \NGen \left(\frac{83 \YtDD}{40}-\frac{39 \YtLL}{40}+\frac{23 \YtUU}{40}\right)
+\ala^2 \alc
+\ala^2 \alb
+\ala \NGen
\nonumber\\
&&
+\alb^2 \NGen \left(\frac{129 \lam}{4}+\frac{63 (\YtD+\YtU)}{16}+\frac{21 \YtL}{16}\right)
+\frac{\ala }{\lam}
+\ala \alc
-1281 \lam^3
\nonumber\\
&&
-\frac{27 \alb^2 }{\lam} \left(\frac{1}{32} \left(\YtD^2+\YtU^2\right)+\frac{\YtD \YtU}{16}\right)
-\frac{20 \alc^2 }{\lam} \NGen (\YtDD+\YtUU)
\nonumber\\
&&
+\frac{\alb^2 \alc }{\lam} \left(27 \z3 (\YtD+\YtU)-\frac{651 (\YtD+\YtU)}{16}\right)
+\frac{407}{160}\frac{ \ala^2 }{\lam} \YtDD \z3
\nonumber\\
&&
+\frac{297}{16\lam} (\YtD \YtLLL+\YtDDD \YtL+\YtL \YtUUU+\YtLLL \YtU)
+\frac{27 \ala^2 \alb }{\lam} \left(\frac{\YtL \z3}{20}+\frac{\YtU \z3}{50}\right)
\nonumber\\
&&
-18 \ala \left(\lam \z3 (\lam+\YtL)+\frac{\lam \YtU \z3}{5}\right)
+81 \ala \left(\frac{1}{40} \left(\YtD^2+\YtU^2\right)+\frac{\YtD \YtU}{20}\right)
\nonumber\\
&&
+\alb \alc \left(\frac{489 (\YtD+\YtU)}{8}-54 \z3 (\YtD+\YtU)\right)
+\alb^2 \alc \NGen \left(18 \z3-\frac{135}{8}\right)
\nonumber\\
&&
-\frac{\alb^3 }{\lam} \NGen \left(\frac{27 (\YtD+\YtU)}{4}+\frac{9 \YtL}{4}\right)
+81 \alb \left(\frac{1}{8} \left(\YtD^2+\YtU^2\right)+\frac{\YtD \YtU}{4}\right)
\nonumber\\
&&
+\frac{\ala^3 \alc }{\lam} \NGen \left(\frac{1683}{2000}-\frac{99 \z3}{125}\right)
+81 \left(\lam \left(\YtD^2+\YtU^2\right)+\frac{\YtLLL \z3}{2}\right)
\nonumber\\
&&
-84 (\YtD \YtLL+\YtDD \YtL+\YtL \YtUU+\YtLL \YtU)
+\frac{\alb^3 \alc }{\lam} \NGen \left(\frac{153}{16}-9 \z3\right)
\nonumber\\
&&
+\ala^2 \alc \NGen \left(\frac{66 \z3}{25}-\frac{99}{40}\right)
+\frac{18 }{\lam} \z3 (\YtDDDD+\YtUDDD+\YtUUUD+\YtUUUU)
\nonumber\\
&&
+\frac{891}{16\lam} (\YtD+\YtU) (\YtDDD+\YtUUU)
-\frac{135}{16\lam} (\YtD+\YtU) (\YtUDD+\YtUUD)
\nonumber\\
&&
+\frac{819}{32}\frac{ \alb^2 }{\lam} \z3 (\YtDD+\YtUU)
+\frac{27}{2} \frac{\alb }{\lam} \z3 (\YtDDD+\YtUUU)
-\frac{9}{16} \frac{\alb^2 }{\lam} \YtL (\YtD+\YtU)
\nonumber\\
&&
-\frac{297}{8} \frac{\alb^3 }{\lam} \z3 (\YtD+\YtU)
-\frac{81}{100}\frac{ \ala^2 \alb }{\lam} \YtD \z3
+\frac{2229}{80}\frac{ \ala \alb }{\lam} \YtUU \z3
+\frac{3}{8}\frac{ \ala \alb }{\lam} \YtD \YtL
\nonumber\\
&&
+\frac{1143}{80}\frac{ \ala \alb }{\lam} \YtLL \z3
+\frac{933}{80}\frac{ \ala \alb }{\lam} \YtDD \z3
-\frac{81}{40}\frac{ \ala \alb^2 }{\lam} \YtU \z3
-\frac{27}{10}\frac{ \ala \alb^2 }{\lam} \YtD \z3
\nonumber\\
&&
-\frac{93}{20}\frac{ \ala \alb }{\lam} \YtUD \z3
+\frac{9}{20}\frac{ \ala \alb^2 }{\lam} \YtL \z3
-\frac{87}{40}\frac{ \ala \alb }{\lam} \YtL \YtU
+\frac{63}{40}\frac{ \ala \alb }{\lam} \YtD \YtU
\nonumber\\
&&
-\frac{70563}{12800}\frac{ \ala^2 \alb }{\lam} \YtU
-\frac{59913}{12800}\frac{ \ala^2 \alb }{\lam} \YtL
-\frac{39627}{12800}\frac{ \ala^2 \alb }{\lam} \YtD
-\frac{48 \alc^2 }{\lam} \left(\YtD^2+\YtU^2\right)
\nonumber\\
&&
-\frac{13437}{256}\frac{ \alb^2 }{\lam} (\YtDD+\YtUU)
-\frac{3411}{64}\frac{ \alb }{\lam} (\YtDDD+\YtUUU)
+\frac{17217}{512}\frac{ \alb^3 }{\lam} (\YtD+\YtU)
\nonumber\\
&&
+\ala^2 \left(\frac{561}{400}-\frac{33 \z3}{25}\right)
-\frac{477}{64}\frac{ \alb }{\lam} (\YtUDD+\YtUUD)
-\frac{12537}{2560}\frac{ \ala \alb^2 }{\lam} \YtD
\nonumber\\
&&
-\frac{36 }{\lam} \left(\YtUD^2+\YtUUDD \z3\right)
-\frac{9309}{2560}\frac{ \ala \alb^2 }{\lam} \YtU
-\frac{5499}{2560}\frac{ \ala \alb^2 }{\lam} \YtL
\nonumber\\
&&
+\frac{457}{3}\frac{ \alc^2 }{\lam} (\YtDD+\YtUU)
-\frac{45}{16}\frac{ }{\lam} \YtL (\YtUDD+\YtUUD)
+\frac{6 }{\lam} \left(\YtLL^2+\YtLLLL \z3\right)
\nonumber\\
&&
-\frac{120 \alc }{\lam} \z3 (\YtDDD+\YtUUU)
-\frac{2957}{800}\frac{ \ala^2 }{\lam} \YtUU \z3
-\frac{2103}{400}\frac{ \ala^2 }{\lam} \YtL \YtU
\nonumber\\
&&
+\frac{24 \alc }{\lam} \z3 (\YtUDD+\YtUUD)
-\frac{16 \alc^2 }{\lam} \z3 (\YtDD+\YtUU)
+\frac{2591}{640}\frac{ \ala \alb }{\lam} \YtDD
\nonumber\\
&&
+\frac{1871}{640}\frac{ \ala \alb }{\lam} \YtUU
-\frac{123}{400}\frac{ \ala^2 }{\lam} \YtD \YtL
+\frac{549}{640}\frac{ \ala \alb }{\lam} \YtLL
+\frac{51}{400}\frac{ \ala^2 }{\lam} \YtD \YtU
\nonumber\\
&&
+\frac{273}{32}\frac{ \alb^2 }{\lam} \YtLL \z3
-\frac{135}{32}\frac{ \ala^2 }{\lam} \YtLL \z3
+\ala \left(\frac{153}{80}-\frac{9 \z3}{5}\right)
-\frac{63}{80}\frac{ \ala \alb }{\lam} \YtL^2
\nonumber\\
&&
-\frac{99}{80}\frac{ \ala \alb }{\lam} \YtU^2
+\frac{81}{80}\frac{ \ala \alb }{\lam} \YtD^2
+\frac{81}{200}\frac{ \ala^3 }{\lam} \YtL \z3
-\frac{711}{16} \alb \lam (\YtD+\YtU)
\nonumber\\
&&
+\frac{459}{8} \alb^2 \z3 (\YtD+\YtU)
-\frac{351}{2} \alb \z3 (\YtDD+\YtUU)
-\frac{301}{64}\frac{ \ala \alb }{\lam} \YtUD
\nonumber\\
&&
-\frac{27}{200}\frac{ \ala^3 }{\lam} \YtD \z3
+\frac{27}{100}\frac{ \ala^3 }{\lam} \YtU \z3
-\frac{117}{4}\frac{ \alb^2 }{\lam} \YtUD \z3
-\frac{99}{10}\frac{ \ala }{\lam} \YtLLL \z3
+\frac{9}{50}\frac{ \ala^2 }{\lam} \YtUD \z3
\nonumber\\
&&
-\frac{51}{10}\frac{ \ala }{\lam} \YtUUU \z3
+\frac{27}{20} \ala \YtL (\YtD+\YtU)
-288 \alc \lam \z3 (\YtD+\YtU)
-\frac{99}{8}\frac{ \alb^3 }{\lam} \YtL \z3
\nonumber\\
&&
+\frac{42}{5}\frac{ \ala }{\lam} \YtUUD \z3
-\frac{39}{5}\frac{ \ala }{\lam} \YtUDD \z3
+\frac{27}{4} \alb \YtL (\YtD+\YtU)
+\frac{15}{2}\frac{ \ala }{\lam} \YtDDD \z3
\nonumber\\
&&
+54 \alb \lam \z3 (\YtD+\YtU)
+\frac{9}{2}\frac{ \alb }{\lam} \YtLLL \z3
-\frac{711}{20} \ala \alb \YtU \z3
-\frac{243}{10} \ala \alb \YtL \z3
\nonumber\\
&&
+\frac{297}{5} \ala \alb \lam \z3
+\frac{249}{16\lam} (\YtUDDD+\YtUUUD)
-\frac{20681}{19200}\frac{ \ala^2 }{\lam} \YtDD
-\frac{18}{5} \ala \alb \YtD \z3
\nonumber\\
&&
-\frac{128829}{64000}\frac{ \ala^3 }{\lam} \YtU
-\frac{106083}{64000}\frac{ \ala^3 }{\lam} \YtL
-\frac{36129}{64000}\frac{ \ala^3 }{\lam} \YtD
-\frac{11269}{3840}\frac{ \ala^2 }{\lam} \YtUU
\nonumber\\
&&
+\frac{54 }{\lam} \left(\YtDD^2+\YtUU^2\right)
-\frac{6669}{1280}\frac{ \ala^2 }{\lam} \YtLL
+\frac{5973}{3200}\frac{ \ala^2 }{\lam} \YtUD
-\frac{39}{4}\frac{ }{\lam} (\YtDDDD+\YtUUUU)
\nonumber\\
&&
-\frac{2241}{800}\frac{ \ala^2 }{\lam} \YtL^2
-\frac{1857}{800}\frac{ \ala^2 }{\lam} \YtU^2
+\frac{7101}{256} \alb^2 (\YtD+\YtU)
-\frac{4479}{256}\frac{ \alb^2 }{\lam} \YtLL
\nonumber\\
&&
+\frac{231}{800}\frac{ \ala^2 }{\lam} \YtD^2
-252 (\YtD+\YtU) (\YtDD+\YtUU)
+\frac{19 \alc }{\lam} (\YtDDD+\YtUUU)
\nonumber\\
&&
+\frac{6453}{32} \alb (\YtDD+\YtUU)
+\frac{5739}{512}\frac{ \alb^3 }{\lam} \YtL
+\frac{567}{128}\frac{ \alb^2 }{\lam} \YtUD
-\frac{5111}{320}\frac{ \ala }{\lam} \YtDDD
\nonumber\\
&&
-\frac{3467}{320}\frac{ \ala }{\lam} \YtUUU
+\frac{2299}{320}\frac{ \ala }{\lam} \YtUDD
-\frac{1337}{320}\frac{ \ala }{\lam} \YtUUD
+\frac{36 }{\lam} \YtLL (\YtDD+\YtUU)
\nonumber\\
&&
-\frac{18 }{\lam} \YtUD (\YtDD+\YtUU)
-\frac{3735}{8} \lam (\YtDD+\YtUU)
+\frac{243}{2} \z3 (\YtDDD+\YtUUU)
\nonumber\\
&&
-\frac{1449}{4} \lam^2 (\YtD+\YtU)
-\frac{1343}{4} \alc (\YtDD+\YtUU)
-\frac{1137}{64}\frac{ \alb }{\lam} \YtLLL
-\frac{3}{32}\frac{ \alb^2 }{\lam} \YtL^2
\nonumber\\
&&
+\frac{29223 \ala \alb \YtU}{640}
+\frac{2247}{200} \ala^2 \YtU \z3
+\frac{2079}{200} \ala^2 \YtL \z3
-270 \lam \z3 (\YtDD+\YtUU)
\nonumber\\
&&
+\frac{15633 \ala \alb \YtL}{640}
+\frac{15459 \ala \alb \YtD}{640}
+\frac{10881 \ala \alb \lam}{160}
+468 \alc \z3 (\YtDD+\YtUU)
\nonumber\\
&&
+\frac{99 }{16\lam} \YtL \YtLLL
-\frac{81}{64}\frac{ \ala }{\lam} \YtLLL
+\frac{177}{200} \ala^2 \YtD \z3
-\frac{96 \alc^2 }{\lam} \YtD \YtU
+306 \alc \lam (\YtD+\YtU)
\nonumber\\
&&
-\frac{63}{4} \YtUD (\YtD+\YtU)
+\frac{567}{50} \ala^2 \lam \z3
+\frac{1599 \ala \lam \YtL}{80}
-\frac{117}{2} \alb \YtLL \z3
\nonumber\\
&&
-\frac{1323}{80} \ala \lam \YtD
+54 \lam \YtL (\YtD+\YtU)
-\frac{531}{10} \ala \YtDD \z3
+\frac{513 \ala \lam \YtU}{80}
\nonumber\\
&&
+\frac{351}{2} \alb^2 \lam \z3
+\frac{243 \ala \YtLL \z3}{10}
-\frac{237}{16} \alb \lam \YtL
+\frac{153}{8} \alb^2 \YtL \z3
-\frac{81}{2} \alb \YtUD \z3
\nonumber\\
&&
+\frac{3 \ala \YtUD \z3}{10}
-\frac{27}{2} \ala \YtUU \z3
+18 \alb \lam \YtL \z3
+18 \ala \lam \YtD \z3
\nonumber\\
&&
+\frac{175399 \ala^2 \YtD}{19200}
+\frac{101791 \ala^2 \YtU}{19200}
+\frac{9435 (\YtDDD+\YtUUU)}{32}
+\frac{43011 \ala^2 \lam}{1600}
\nonumber\\
&&
-\frac{183}{32} (\YtUDD+\YtUUD)
-\frac{369 \ala^2 \YtL}{6400}
+\frac{2367 \alb^2 \YtL}{256}
-54 \z3 (\YtUDD+\YtUUD)
\nonumber\\
&&
+\frac{7949 \ala \YtDD}{160}
-\frac{375}{8\lam} \YtUUDD
+\frac{3531 \alb^2 \lam}{64}
+\frac{3353 \ala \YtUU}{160}
+\frac{108 }{\lam} \YtDD \YtUU
\nonumber\\
&&
+\frac{411 \ala \lam^2}{10}
-\frac{33}{4\lam} \YtUDUD
+\frac{2151 \alb \YtLL}{32}
-\frac{13}{4\lam} \YtLLLL
+\frac{495 \alb \YtUD}{16}
-\frac{483 \lam^2 \YtL}{4}
\nonumber\\
&&
-\frac{441 \ala \YtLL}{32}
+\frac{411 \alb \lam^2}{2}
-\frac{1245 \lam \YtLL}{8}
+\frac{121 \ala \YtUD}{16}
-\frac{1197 \lam \YtUD}{4},
\label{eq:anom_v_3}
\end{eqnarray}
}
where the following notation was used for the gauge couplings:
\begin{equation}
	 a_i   =   \left(\frac{5}{3} \frac{g_1^2}{16\pi^2}, \frac{g_2^2}{16\pi^2}, \frac{g_s^2}{16\pi^2}\right).
	\label{eq:coupl_notations}
\end{equation}

Let us also mention that the results for $\beta_{\lam}$, $\gamma_{m^2}$ and $\gamma_{v}$ are independent of gauge-fixing parameters and
the corresponding renormalization constants satisfy the so-called pole equations \cite{'tHooft:1972fi}. 
This serves as a crucial test of the correctness of the calculated three-loop contributions.

\begin{figure}[th!]
	\centering{\includegraphics[width=\textwidth]{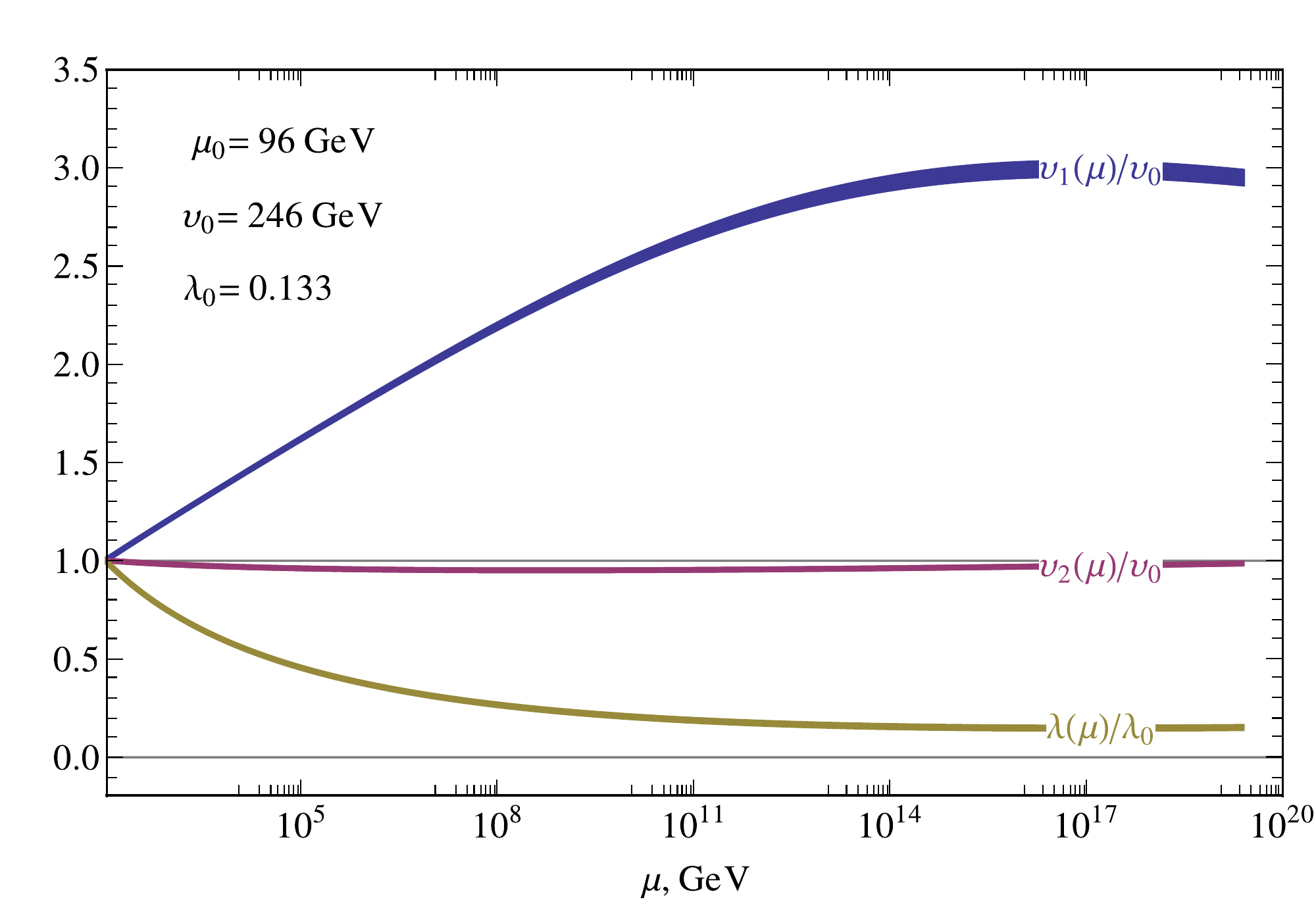}}
	\caption{The scale dependence of $v_1(\mu)\equiv\sqrt{m^2(\mu)/\lambda(\mu)}$ and $v_2(\mu)$, which minimizes the SM effective potential.
	The width of the curves corresponds to the difference between two- and three-loop running. 
All parameters are normalized by their initial values at the scale $\mu_0$ and
	we assume that $v_1(\mu_0) = v_2(\mu_0) = v_0.$. The arrow points to the scale at which $\beta_\lambda = 0$.
	}
	\label{fig:1}
\end{figure}
\begin{figure}[ht!]
	\centering{\includegraphics[width=\textwidth]{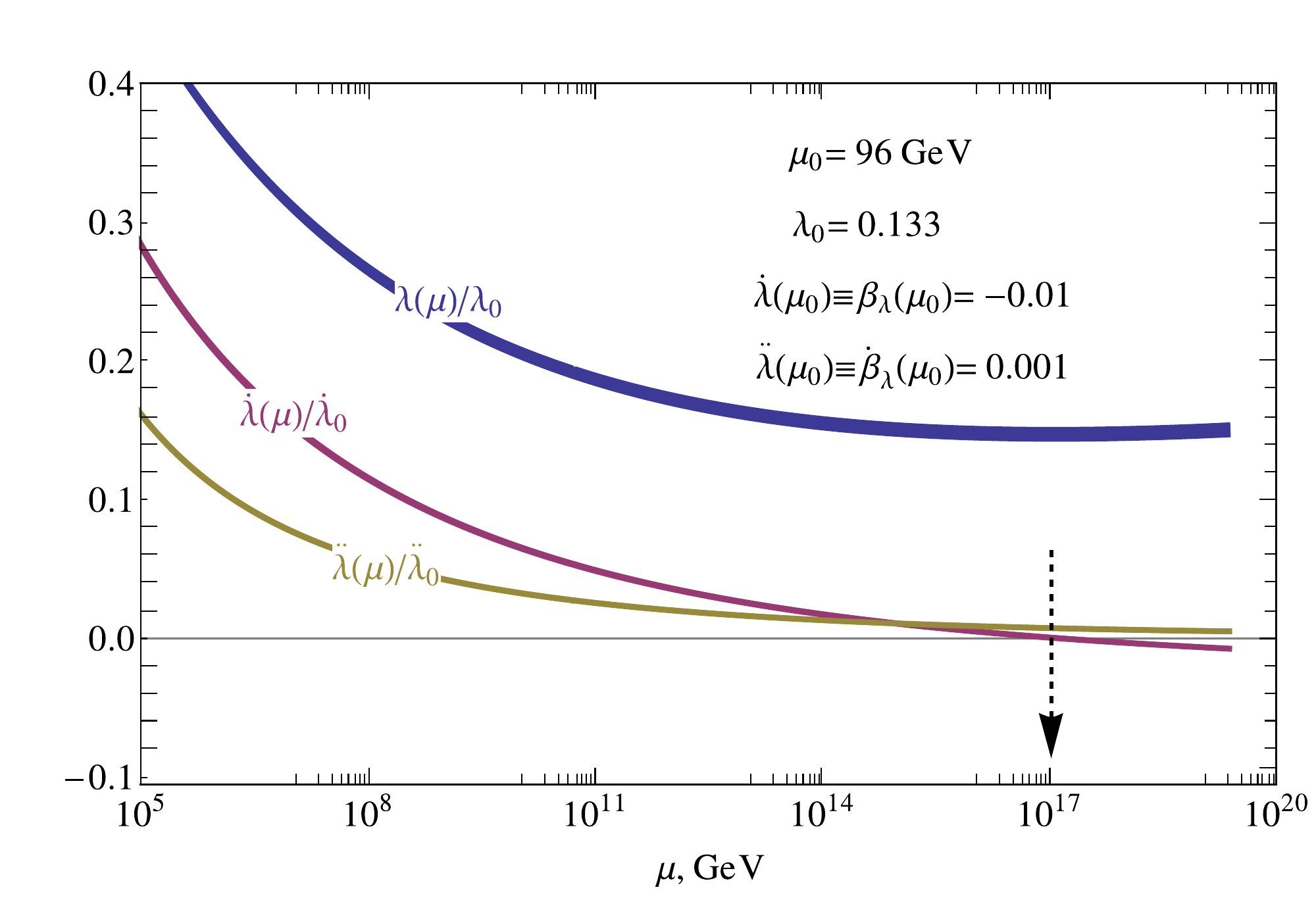}}
	\caption{The running of the Higgs self-coupling is given together 
		with the scale dependence of its first $(\beta_\lambda)$ and second derivatives. 
	The width of the curves corresponds to the difference between two- and three-loop running. 
	All parameters are normalized by their initial values at the scale $\mu_0$. The arrow points to the scale at which $\beta_\lambda = 0$.}
	\label{fig:2}
\end{figure}

It is worth pointing that the expressions \eqref{eq:anom_v_1}-\eqref{eq:anom_v_3} do not coincide with the anomalous dimension\footnote{The corresponding expression can also be found in
ancillary files of the arXiv version of the paper.} of the Higgs doublet 
$\gamma_\Phi = -\frac{1}{2} \frac{d \ln Z_h}{d\ln\mu^2}$ which in the Landau gauge coincide with the anomalous dimension of the
VEV obtained via minimization of the effective potential (see Refs.~\cite{Ford:1992pn,Martin:2001vx,Sperling:2013eva})

In Fig.~\ref{fig:1} one can see an example of the VEV running driven by two different anomalous dimensions: $v_1(\mu)$ by $\gamma_v$ from Eqs.~\eqref{eq:anom_v_1}-\eqref{eq:anom_v_3} and
$v_2(\mu)$ by $\gamma_{\Phi}$ in the Landau gauge. 
The initial scale is chosen to be $\mu_{0} \simeq 96$~GeV~\cite{Jegerlehner:2013dpa} at which one expects the threshold corrections for $v_1(\mu)$ 
to be small. The boundary values for the couplings are also taken from Ref.~\cite{Jegerlehner:2013cta} and we made the assumption that $v_2(\mu_0) = v_1(\mu_0) = v_0 \simeq 246$~GeV.  
For convenience, we divide all the running quantities in Fig.~\ref{fig:1} by
their boundary values. 
It is clear that $v_1(\mu)$ increases significantly with $\mu$, while the scale dependence of $v_2(\mu)$ is rather smooth.
This is due to the fact that the anomalous dimension of $v_1(\mu)$ is correlated with $-\beta_\lambda/\lambda$, and at $\mu_0$ we have a large positive contribution
$-\beta_{\lambda}(\mu_0)/\lambda_0 \simeq 0.08$ to  $\gamma_v$.
In Fig.~\ref{fig:2}  the scale dependence of $\lambda(\mu)$ and $\beta_\lambda(\mu)$ is presented. 
In addition, we plot the second derivate $\ddot{\lambda}(\mu)$, which can be of some interest in scenaria
with  $\lambda=\beta_\lambda = 0$ at some scale.
Form Fig.~\ref{fig:2} one can see that for a chosen set of initial parameters~\cite{Jegerlehner:2013cta} the beta-function $\beta_\lambda$ reaches zero at $10^{17}$~GeV,
while $\lambda$ and $\ddot{\lambda}$  are still positive at this scale.

It is fair to mention that different implementation~\cite{Buttazzo:2013uya} of threshold corrections, which relate the $\MS$ parameters 
to some measured quantities, leads to a different boundary value of the top Yukawa coupling. 
The latter drives $\lambda$ to negative values at the scales of order $10^{10}$ GeV rendering the SM vacuum unstable.
Since, in our opinion, both procedures\footnote{The essential difference in matching procedures of 
	Refs.~\cite{Jegerlehner:2013cta}~and~\cite{Buttazzo:2013uya} stems from the way one treats the so-called tadpole contributions \cite{Fleischer:1980ub} or, in other words, 
	whether $v_1(\mu)$ or $v_2(\mu)$ is used in the relations between \MS-running masses and couplings.}, if implemented consistently, 
should render the same values for dimensionless couplings, this discrepancy requires further investigation.

To conclude, by explicit calculation we extended our results presented in Ref.~\cite{Bednyakov:2013eba} to the case of complex Yukawa matrices.
We also provided the anomalous dimensions $\gamma_\Phi$ and $\gamma_v$ of the Higgs doublet and the running vacuum expectation value (VEV) defined as $v \equiv \sqrt{m^2/\lambda}$, 
respectively. In addition, the scale dependence of the considered quantities are studied numerically.

\subsection*{Acknowledgments}
The authors would like to thank M.~Kalmykov for very stimulating discussions and interesting suggestions, as well as
F.~Jegerlehner, for multiple comments.
AVB is grateful to M.~Kompaniets for fruitful discussions of graph isomorphism and his 
help with the \texttt{graph\_state} library.
This work is partially supported by RFBR grants 11-02-01177-a, 12-02-00412-a, RSGSS-4801.2012.2, JINR Grant~No.~13-302-03.
The research of V.N.~Velizhanin is supported by a Marie Curie International Incoming Fellowship within the 
7th European Community Framework Programme, grant number PIIF-GA-2012-331484.

\end{document}